# Tiny-PPG: A Lightweight Deep Neural Network for Real-Time Detection of Motion Artifacts in Photoplethysmogram Signals on Edge Devices

Yali Zheng[#1], Chen Wu[#1,] Peizheng Cai[1], Zhiqiang Zhong[1], Hongda Huang[1] and Yuqi Jiang[2]

[1] Department of Biomedical Engineering, College of Health Science and Environmental Engineering, Shenzhen Technology University, Shenzhen, China.

[2] Department of Surgery, Chinese University of Hong Kong, Hong Kong, China.

# These authors are equally contributed. *Corresponding author: zhengyali@sztu.edu.cn .

**Abstract**

Photoplethysmogram (PPG) signals are easily contaminated by motion artifacts in real-world settings, despite their widespread use in Internet-of-Things (IoT) based wearable and smart health devices for cardiovascular health monitoring. This study proposed a lightweight deep neural network, called Tiny-PPG, for accurate and real-time PPG artifact segmentation on IoT edge devices. The model was trained and tested on a public dataset, PPG DaLiA, which featured complex artifacts with diverse lengths and morphologies during various daily activities of 15 subjects using a watch-type device (Empatica E4). The model structure, training method and loss function were specifically designed to balance detection accuracy and speed for real-time PPG artifact detection in resource-constrained embedded devices. To optimize the model size and capability in multi-scale feature representation, the model employed depth-wise separable convolution and atrous spatial pyramid pooling modules, respectively. Additionally, the contrastive loss was also utilized to further optimize the feature embeddings. With additional model pruning, Tiny-PPG achieved state-of-the-art detection accuracy of 87.4% while only having 19,726 model parameters (0.15 megabytes), and was successfully deployed on an STM32 embedded system for real-time PPG artifact detection. Therefore, this study provides an effective solution for resource-constraint IoT smart health devices in PPG artifact detection.

## 1. Introduction

Cardiovascular diseases have emerged as the leading cause of mortality in the modern society. With the advancement of sensing, electronic, and information technologies, smart wearable medical devices such as watches, glasses, and clothing have been proposed for the unobtrusive measurement of vital signs like electrocardiogram (ECG) or photoplethysmogram (PPG) [1] to monitor cardiovascular health. These devices hold significant potential in the prevention, precise diagnosis, and treatment of cardiovascular diseases.

PPG is a simple, non-invasive, and cost-effective optical technique that measures blood volume changes in microvasculature and has been widely employed to estimate physiological parameters such as heart rate, blood pressure [2][3], oxygen saturation [4], and respiratory rate [5], etc. However, PPG signals are often interfered with by various noises, especially motion artifacts (MA) caused by relative movements between the sensor and skin. MAs are difficult to remove because they overlap with the signal frequency [6]. Early studies attempted to segment PPG signal into individual pulses using the derivative of the signal and then evaluated the signal quality by template matching [8]. To cope with the nonlinear and nonstationary changes in the pulse morphology caused by physiological factors, dynamic time warping method was introduced to segment PPG pulses [9]. However, pulse segmentation was still very challenging in complex motion scenarios. Therefore, research began to divide PPG signals into segments with fixed window size (or sliding window) and assess the signal quality of each segment by diverse methods. Some employed statistical metrics like skewness, kurtosis, and entropy [10]. More prevalent approaches relied on feature engineering, entailing the extraction of handcrafted features from the PPG signal across various domains, including time, frequency, wavelet, and complexity. These features were then utilized in supervised machine learning models for classification, including hierarchical decision rules [11], SVM [12,13], decision tree [7], Bayesian method [14], multilayer perceptron neural network [12], as well as unsupervised methods like the elliptical envelope algorithm [15], and semi-supervised methods [16]. Among these models, there have been some energy-efficient methods targeting for on-board motion artifact detection on resource-constrained Internet of Things (IoT) devices [15, 16].

Recent studies have utilized deep learning methods in the application of PPG artifact detection [17,18]. For example, Goh et al. proposed a 1-D Convolutional Neural Network (CNN) that divided the PPG signal into 5-second segments, and was applied to ward data from PhysioNet and a private dataset collected during short-term activities, achieving an accuracy of $92 \pm 2\%$ [17]. Chen et al. developed a semi-supervised learning artifact detection approach using partially labeled data [19]. Zargari et al. achieved 99% accuracy with a 1-D CNN by adding artificially generated noises [20]. Chen J et al. [21] transformed the PPG signal into a 2-D time-spectrogram image by short-time Fourier transform and used 2-D CNN for classification, achieving 98% accuracy on static data from general wards. Liu S H et al. and X. Liu et al. also transformed PPG signals into 2-D images and employed a 2-D CNN model to attain $93 \pm 3\%$ accuracy on PPG data collected from wards and short-term running scenarios [16].

Despite achieving high accuracy, these deep learning methods [17, 18, 20, 21] have not undergone thorough evaluation under daily-life activities, where the morphologies of PPG artifacts are more complex. Furthermore, these methods commonly employ fixed window sizes for signal segmentation, so the models make decisions on whether a signal segment is clean or contains artifacts. This approach may lead to inaccuracies, as PPG artifacts in real-life conditions can have diverse lengths, resulting in partial artifact submergence within the signal. To address the limited temporal resolution of these methods, [24] recently proposed an Unet-based semantic *segmentation* model for PPG artifact detection. Instead of using fixed window sizes, segmentation models classify each sample point of the signal as clean or containing artifacts, enabling more precise detection. In datasets encompassing a wide range of daily activities, this state-of-the-art model achieved a segmentation accuracy of 87.3% [24]. Compared to the early template matching methods and the recent feature engineering methods, the deep segmentation models can perform motion artifact detection with high temporal resolution.

However, the model complexity and size of existing deep learning models for PPG artifact detection were too large to be implemented on IoT devices. Furthermore, as wireless transmission is more energy-intensive than on-board computations, it is highly desirable to develop on-board PPG artifact detection methods that can detect artifacts in real-time and remove these uninformative data before transmission to save energy. Fortunately, there have been some research on pruning neural networks and deploying them on edge devices for other healthcare applications [25], such as the Tiny-HR for heart rate estimation and the ANNet for ECG anomaly detection [26,27].

In this study, we developed a lightweight model called Tiny-PPG to provide a precisive and efficient solution for PPG artifact detection in IoT-based wearable and smart health devices. It is based on the deep segmentation model, which can classify each sample point of the signal as clean or containing artifacts, enabling more precise detection. Additionally, by employing depth-wise separable convolution (DSC) [29] and Atrous Spatial Pyramid Pooling (ASPP) [30], Tiny-PPG can achieve high accuracy with a small model size. We further improved the detection accuracy by incorporating contrastive learning, and reducing parameter quantity through model pruning. Compared to existing PPG artifact detection methods, Tiny-PPG demonstrated state-of-the-art accuracy in artifact detection with a very small network, making it easy to deploy on embedded devices for real-time PPG artifact segmentation. The contributions of this study can be summarized as follows:

(1) A lightweight deep neural network Tiny-PPG was proposed. Through validation on a dataset under daily activities, the model achieved state-of-the art accuracy.

(2) Tiny-PPG model was optimized for resource-constraint embedded devices, and successfully deployed on an STM32 embedded device for real-time PPG artifact detection.

(3) The effectiveness of incorporating contrastive loss function in the task of PPG artifact segmentation was investigated, which was determined by several factors especially the ingenious design of positive and negative pairs.

## 2. Method

### 2.1 Dataset and Preprocessing

The PPG-DaLiA public dataset used in this study was from two references [17][21]. The dataset consisted of PPG signals collected from 15 subjects (8 female and 7 male) aged 21-55 years old using a wrist-worn device (Empatica E4) at a sampling rate of 64 Hz. Each subject was asked to perform various daily activities for about 2 hours, including low-intensity activities such as sitting, driving, lunch break, working, as well as medium-intensity activities such as walking, ascending and descending stairs, cycling, and high-intensity arm movements such as table soccer.

PPG signal was first filtered by a 0.9-5 Hz band-pass filter to remove baseline drift, power frequency interference and other noise. Next, the PPG signals were divided into 30-sec segments, resulting in a total of 4,305 segments. Min-Max normalization was then applied to each segment to normalized the amplitude to the range of [0, 1].

The fine-grained segmentation labels were created by the authors of the reference [24]: they developed a web-based tool which enables the annotator to accurately select segments of the signal and designate them as artifacts, and then automatically transcribes the users' annotations into binary segmentation labels.

### 2.2 Study Pipeline and Modeling

Figure 1 presents an overview of the study pipeline, consisting of three stages: contrastive training of Tiny-PPG, pruning of Tiny-PPG, and deployment and inference on STM32 embedded system. Firstly, a novel lightweight deep convolutional neural network, called Tiny-PPG, was designed, incorporating Depthwise Separable Convolution (DSC) [29] and Atrous Spatial Pyramid Pooling (ASPP) [30] modules to achieve high segmentation accuracy with a small model size. In addition, a contrastive learning loss was introduced during the training process to learn contrastive representations that bring intra-class samples closer and inter-class samples farther apart. In the second stage, redundant connections of the model were pruned to further decrease the model size. Finally, the model was deployed on an STM32 embedded device, allowing for real-time PPG artifact segmentation.

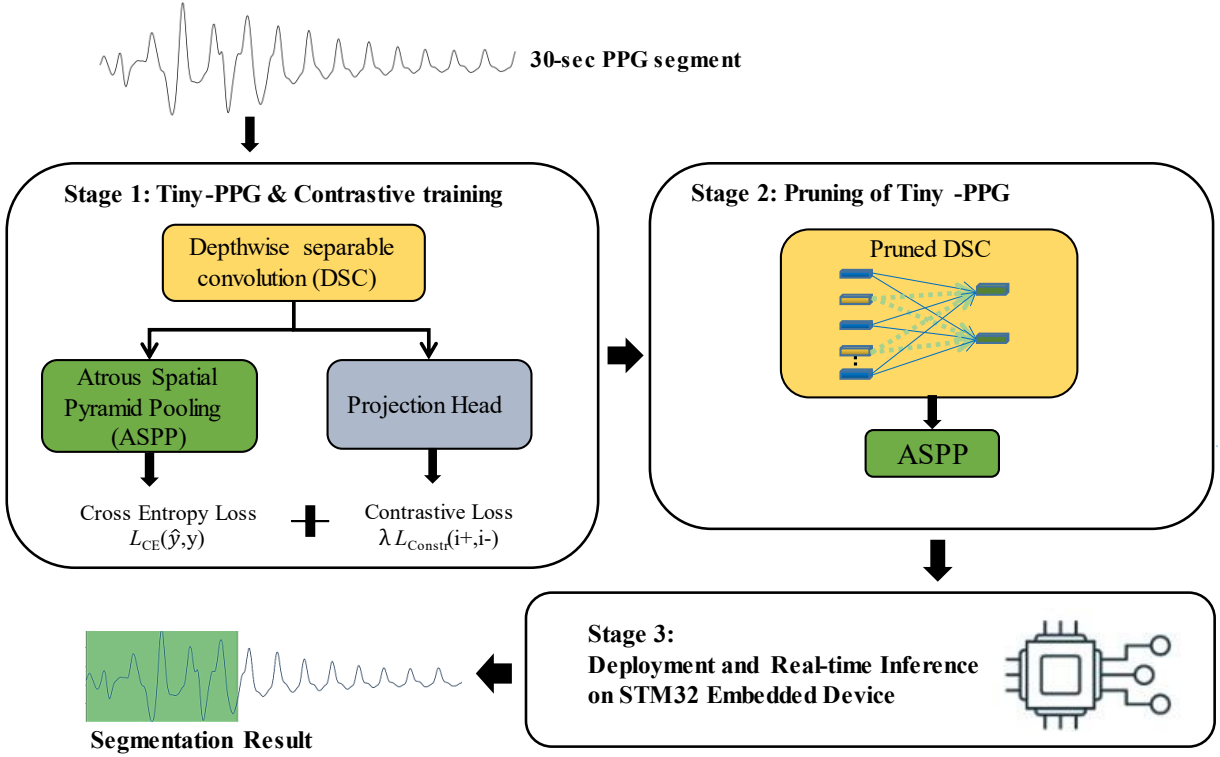

Figure 1. The overview of the study pipeline

*2.2.1 Model Architecture*

The model architecture of Tiny-PPG is shown in Figure 2. It is a 1-D fully convolutional network that utilizes two key modules, the DSC [29] and ASPP [30], to achieve high accuracy in PPG artifact detection with a small number of parameters. Unlike standard convolution, DSC employs two separate convolutional steps: the depthwise convolution, which captures temporal correlations within each channel, and the pointwise convolution, which captures correlations across channels. This approach significantly reduces the number of parameters without compromising accuracy [29]. Furthermore, the atrous convolution, which upsamples the feature vector by convolution kernels spaced with zeros, can expand the receptive field of the features without increasing the number of parameters. By employing multiple atrous convolution kernels with different atrous rates, ASPP, a pyramid-like atrous convolution [30], provides multi-scale features, thereby enabling the detection of artifacts of various scales. The detailed structure of the model is described below:

1) **Depth-wise separable convolution**. The 30-sec PPG segment was first passed through four DSC modules. Each DSC module consists of four layers with the exception of the last module, which does not contain a maxpooling layer, and the four layers are: a depthwise convolution (Depthwise_conv), a pointwise convolution (Pointwise_conv), a batch normalization (BN) and a maxpooling layer with ReLU activation function. The purpose of the BN and ReLu layers is to prevent the gradients from vanishing or exploding. The DSC modules employed different numbers of input and output channels, and sizes of kernels in each 1-D convolutional layer, specifically: (1, 32, 80), (32, 64, 40), (64, 128, 20), and (128, 512, 7), respectively.

2) **Atrous Spatial Pyramid Pooling**. After the PPG segment was processed by the DSC modules for feature extraction, four channels of atrous convolution with the same kernel size of 3 and different atrous ratios

(4, 8, 12, and 16) were used to expand the receptive fields and provide multi-scale features. The output features from the four channels were then concatenated to form a tensor of size (4, 240), which was subsequently upsampled and convolved to the final feature vector of size (1, 1920). Finally, the model output the final segmentation result of the same length as the input signal using the sigmoid activation function.

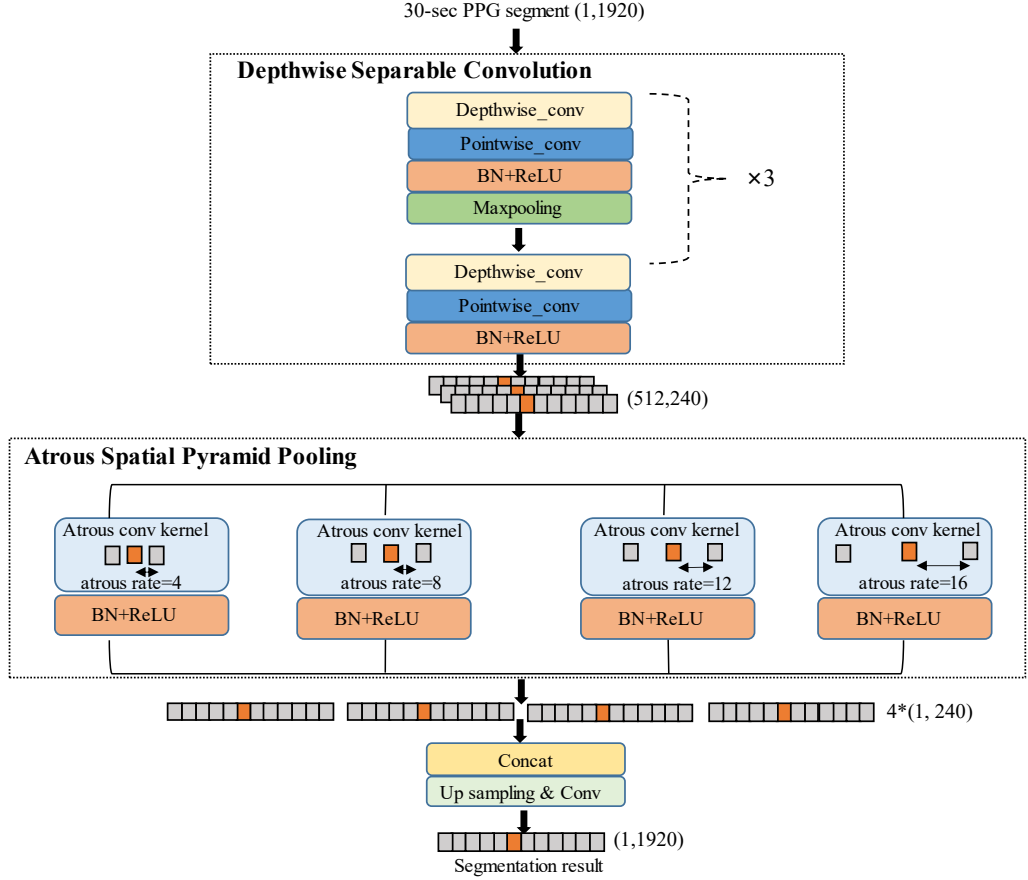

Figure 2. The model architecture of Tiny-PPG

### 2.2.2 Contrastive learning of Tiny-PPG

The traditional loss function only penalizes individual predictions without considering the relationships among different samples. To address this limitation, contrastive learning loss is proposed as a metric loss function to regularize the feature space by increasing the similarities between intra-class samples and decreasing the similarities between inter-class samples in a projected embedding space. By incorporating contrastive learning loss into PPG artifact segmentation, it is assumed that samples within the same class (either clean or artifact) are likely to be more similar than those from different classes (clean vs artifact), which can guide Tiny-PPG to learn better feature representations that improve the intra-class compactness and inter-class dispersion to achieve better discrimination between the two classes.

The training process of Tiny-PPG is shown in Figure 3. The overall objective loss function for Tiny-PPG training is shown in formula (1), including the traditional cross-entropy loss $L_{CE}$ and the contrastive loss $L_{contrast}$

$$L = \sum_i (L_{CE}(\hat{y}_i, y_i) + \lambda * L_{Contrast}(i^+, i^-)) \qquad (1)$$

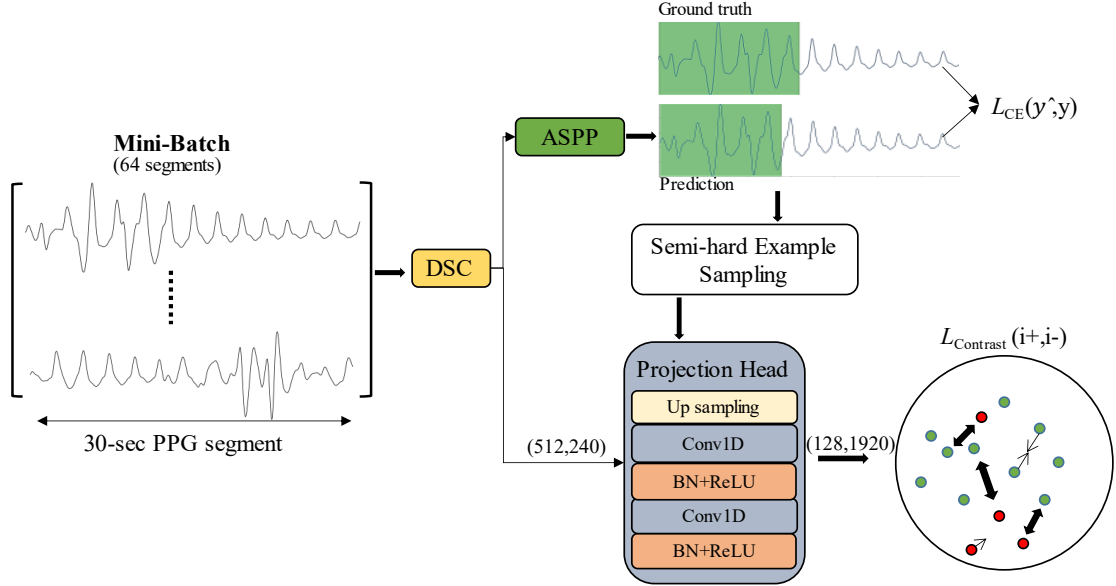

Figure 3. The contrastive training process of Tiny-PPG model for PPG artifact detection

where λ represent the weighting coefficient of contrastive loss in the total loss and was set to 0.1 in this study. To compute the contrastive learning loss, a projection head was added after the DSC module to map the DSC embeddings to another embedding space where sample distances could be calculated. The project head consists of a upsampling layer and two 1-D convolutional layers with BN and ReLU activation. The projection head was only used in the training process and was removed during inference. The contrastive learning loss is shown in formula (2),

$$L_{\text{Contrast}}(i^+, i^-) = \frac{1}{|P_i|} \sum_{i^+ \in p_i} -\log \frac{\exp\left(\frac{i \cdot i^+}{\tau}\right)}{\exp\left(\frac{i \cdot i^+}{\tau}\right) + \sum_{i^- \in N_i} \exp\left(\frac{i \cdot i^-}{\tau}\right)} \quad (2)$$

where $P_i$ and $N_i$ represent the sample embedding collections of positive ($i^+$) and negative ($i^-$) samples of an anchor sample $i$, respectively. "·" denotes inner product, and τ>0 is the temperature hyper-parameter.

**Semi-hard example sampling strategy.** The contrastive loss was computed using a semi-hard example sampling strategy. In each mini-batch, 200 anchor points (their embeddings) were selected from each signal segment for the calculation of the loss. The 200 points were not necessarily from the same signal segment. 50% of them are from easy examples, meaning they were correctly predicted by the model according to $L_{\text{CE}}$, while the other 50% were hard examples, meaning they were incorrectly predicted. Of the hard examples, half of them (i.e., 100 samples) were clean samples wrongly predicted as artifacts, and the remaining half were artifact samples wrongly predicted as clean. If there were fewer than 100 hard or easy examples in the mini-batch, all of them were selected, and the remaining examples were randomly sampled from the mini-batch to reach a total of 200.

**Memory bank.** To enable the learning of contrastive features from more samples across mini-batches, a memory bank was utilized to store the embeddings of samples from other mini-batches. The memory bank had a length of 1000 for both artifact and clean categories. In each mini-batch, the embeddings of 50 anchor points were randomly selected and added to the memory bank. If the number of embeddings in the memory bank exceeded its length, the oldest embeddings were removed to make space for the new ones. The embeddings in the memory bank were only used as positive and negative samples, and not as anchor samples for other mini-batches. Therefore, contrastive learning was performed within 1000 positive and/or negative pairs for the anchor points from each mini-batch.

*2.2.3 Model Pruning*

To optimize the Tiny-PPG model for resource-constrained embedded devices [31], structural pruning was applied to remove unimportant convolutional channels to further reduce the model size. Specifically, the pointwise convolutional layer in the DSC block was pruned using channel scaling factors of each BN layer. These factors were used to weigh the importance of different convolution channels and ranked based on $L1$ regularization as a sparse penalty to identify the most important channels. The pruning ratio k% was calculated by dividing the number of pruned channels by the number of original channels in the entire network. The corresponding weights of the identified channels with less importance were then set to zero to achieve a desired pruning ratio. The pruned model parameters were fine-tuned after pruning to ensure that the model maintained its performance. The overall goal of model pruning was to reduce the model size while maintaining or even improving its accuracy. The pruning process is illustrated in Figure 4.

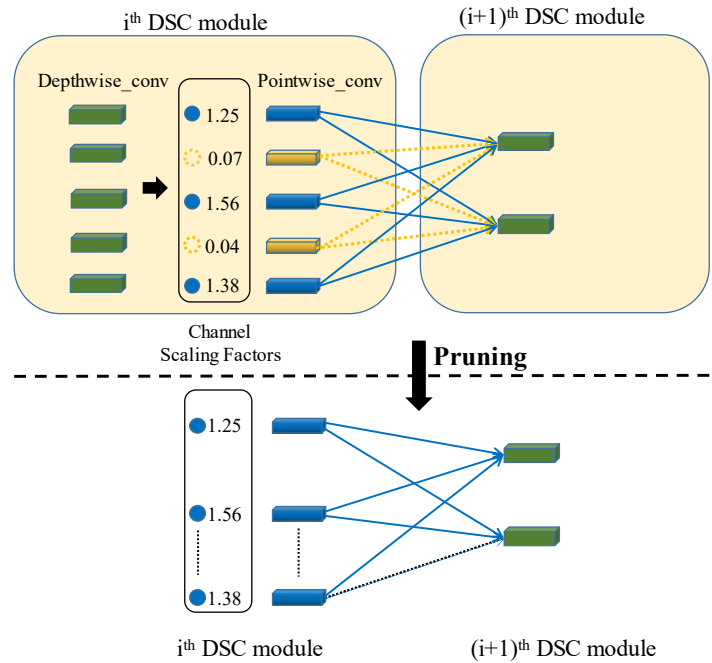

Figure 4. Pruning of unimportant convolutional channels: *i*-th represents the convolution layer of the *i*-th layer and (*i+1*)-th represents the convolutional layer of the next layer.

*2.2.4 Implementation on STM32 Embedded System*

A STM32 development board was utilized for deploying the models for real time inference, which is equipped with a STM32F746NGH6 MCU featuring an ARM 32-bit single-core Cortex-M7 processor and FPU, operating at up to 216 MHz, and 512 kilobytes RAM and 1megabyte built-in flash memory to store the model parameters and input data. Real-time signal and segmentation results were displayed on a 480×272 LCD screen on the board. The PPG signal was acquired using a pulse sensor module with a green LED and photodetector (World Famous Electronics, IIC), and was filtered by an analog band-pass filter on the module between 0.054~1.6Hz. STM32Cube.AI was utilized to convert the pre-trained and pruned model into optimized C code for STM32 microcontrollers within the integrated development environment STM32CubeIDE. As the model is already small, further compression was not performed on the board. Five lightweight models in Table 1 that fit the memory footprint was deployed on the embedded system, including the 1D-CNN with 3-sec sliding window [17], the baseline FCN, the baseline FCN-ASPP, the Tiny-PPG and pruned Tiny-PPG.

## 2.3 Performance Evaluation

We adopted the data division scheme suggested in reference [24]. Briefly, the dataset was partitioned into training and testing subsets. Then ten-fold cross-validation was conducted on the training subset. In each fold, 10% of the training data was randomly chosen for validation. The ten models resulting from the ten-fold cross-validation were each evaluated on the testing set. The arithmetic average and standard deviation (SD) of the DICE scores across these ten models are presented in Table 1. The model was built in Pytorch 1.9.0 using a NVIDIA Tesla V100 PCLE graphics card with 32 GB Video RAM. The hyperparameters used during training included a batch size of 64, an initial learning rate of 0.0005, and an iteration number of 1000 epochs. The learning rate was decreased by 50% every 100 epochs, and the optimizer used was Adam. The performance of the model was evaluated using the DICE coefficient, which can be calculated from True Positive (TP), False Positive (FP) and False Negative (FN), i.e.,

$$\text{DICE} = \frac{2\text{TP}}{2\text{TP} + \text{FP} + \text{FN}} \tag{3}$$

where the artifact sample point is considered as positive and clean one as negative. In addition, number of model parameters, model size and multiply-accumulate operations (MACs) were also summarized in Table 1 to evaluate the potential of the models to be deployed on embedded systems for real-time artifact detection. Two functions, try_count_flops() or profile(), were employed to calculate the MACs, depending on whether the model was developed in Tensorflow or Pytorch. The model size is the file size of each model.

Several previous models for PPG motion artifact detection were implemented and validated with the same dataset for performance comparison with Tiny-PPG, including the pulse segmentation template method [34], two 1-D CNN models [17][18] and U-net [24]. The advanced deep learning models such as DeeplabV3 [32] and Deconvnet [33] were also implemented. Ablation experiments were performed to evaluate the contribution of DSC and ASPP modules in performance gain of the Tiny-PPG, including a baseline fully convolution network (*baseline FCN*) by removing the ASPP module and replacing the depthwise_conv and pointwise_conv layers of Tiny-PPG with standard convolutional layers, and an FCN with ASPP module (*baseline FCN-ASPP*). The influence of different memory bank configurations in contrastive training were also assessed. Table 1 provides a concise summary of the DICE coefficients for all models, along with the number of model parameters, model size, and MACs for all deep learning models.

## 3. Results

**Model architecture & pruning**. As shown in Table 1, compared to the pulse segmentation template matching technique and the sliding-window based CNN models, the segmentation models such as Unet, DeeplabV3 and Deconvnet demonstrated superior performance on the dataset with daily-life activities. However, it is important to acknowledge that these models came with considerable size expansions. Remarkably, our *baseline FCN* model showcased a performance level similar to the state-of-the-art U-net model [24] yet with a significantly lower parameter amount. The integration of the ASPP module into the baseline FCN (referred to as *baseline FCN-ASPP*) further improved the DICE coefficient by 0.7%, with only a slight increase in the number of parameters and MAC. With the substitution of conventional convolutional layers with DSC modules in the *baseline FCN-ASPP* model, the resulting *Tiny-PPG* model realized a substantial 55% reduction in parameters, reaching around 86k. This model reduction hardly impacted the DICE coefficient, merely decreasing from 87.7% to 87.4%. Additionally, the MAC was reduced to around 10% of the original value. Through pruning and fine-tuning, the number of model parameters in *Tiny-PPG* was significantly reduced to 19k, with only 17M of MAC. This came at the expense of only a marginal reduction in the DICE coefficient from 87.7% to 87.4%. Consequently, the proposed *Tiny-PPG* model reduced the number of parameters while achieving high detection accuracy compared to existing models. Figure 5 shows a typical example of the segmentation results of different models.

Table 1 Performance comparison of the proposed Tiny-PPG with other models

| Models | DICE | Number of Parameters | Model Size (Mega Byte, MB) | Number of Multiply-accumulate operations (MACs) |
| --- | --- | --- | --- | --- |
| Pulse segmentation template matching [34] | 69.7% ± 3.2% | NA | NA | NA |
| Deep CNN with 3-sec sliding window [18] | 85.0% ± 1.0% | 641,185 | 16.0 MB | 36,990,208 |
| 1D-CNN with 3-sec sliding window [17] | 80.7% ± 0.1% | 47,617 | 0.62 MB | 19,332,096 |
| U-net [24] | 87.3 ± 0.18% | 2,087,281 | 27.6 MB | 2,750,798,880 |
| DeeplabV3 [32] | 86.4 ± 0.12% | 4,002,480 | 47.6 MB | 3,466,561,935 |
| Deconvnet [33] | 86.0 ± 0.27% | 2,065,345 | 24.4 MB | 2,310,460,210 |
| Baseline FCN | 87.0 ± 0.11% | 179,505 | 0.77 MB | 549,526,272 |
| Baseline FCN-ASPP | 87.7 ± 0.14% | 182,351 | 0.77 MB | 550,023,552 |
| Tiny-PPG | 87.4 ± 0.23% | **85,949** | **0.36 MB** | **53,251,936** |
| *Pruned* Tiny-PPG | 87.2 ± 0.16% | **19,726** | **0.15 MB** | **16,961,870** |

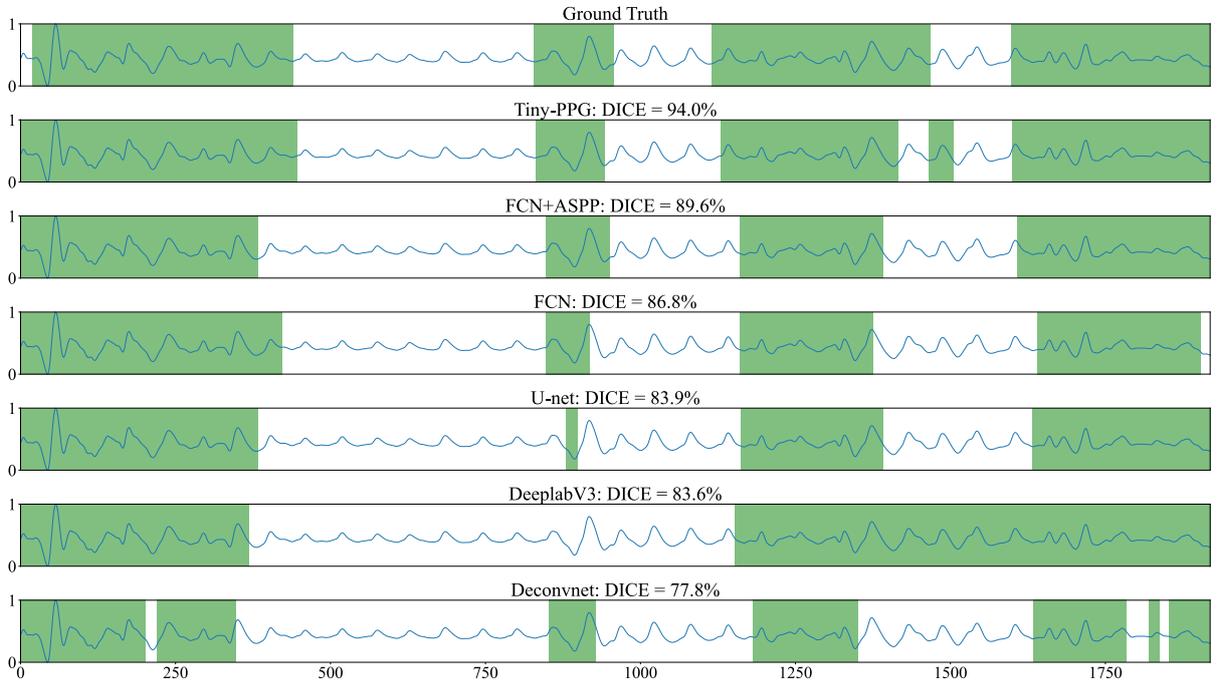

Figure 5. A typical example of the artifact segmentation results of our pruned Tiny-PPG and other models on a 30-sec PPG signal segment. The green area indicates the ground truth and segmented artifacts.

As shown in Figure 6, pruning without finetuning led to significant drop on the segmentation accuracy, and the performance degraded with the pruning ratio. The accuracy can be reserved through fine-tuning. It is also found that, the pruning was focused on the last DSC module, of which the number of DSC channels was reduced by more than 50% (from 512 to 158). The DICE achieved the optimal at the pruning ratio of 50%. This suggests that there is a considerable amount of parameter redundancy in existing convolutional neural networks and pruning can effectively reduce the redundancy.

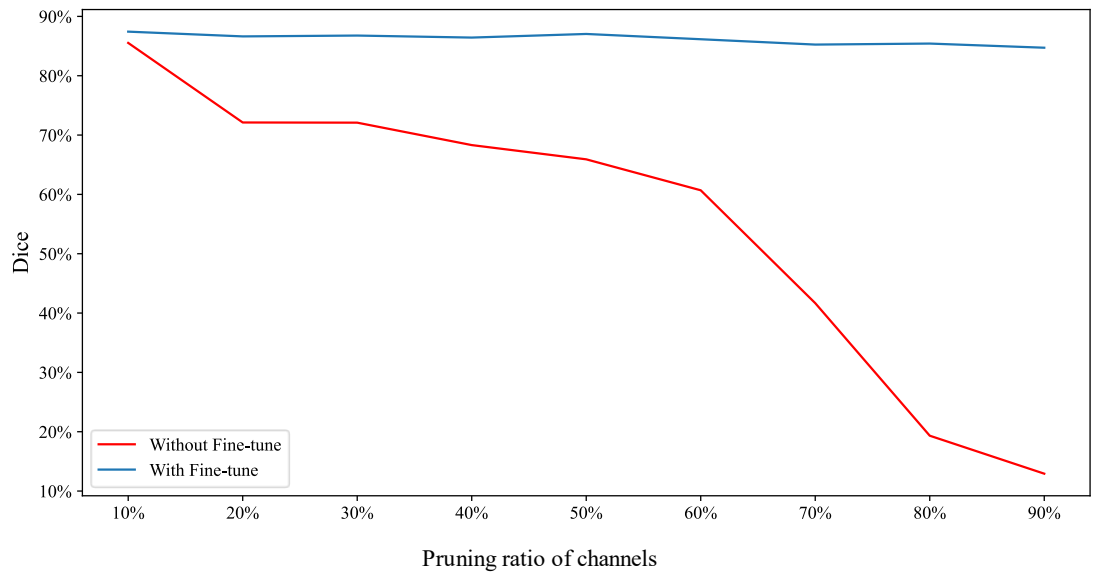

Figure 6. The effect of pruning on segmentation accuracy of Tiny-PPG with (blue) and without (red) finetuning

**Contrastive Learning**. Table 2 presents the results of the ablation experiments conducted to evaluate the effects of contrastive learning with different configurations. The results show that separating inter-class samples and aggregating intra-class samples (i.e., Artifact & Clean) achieved a better DICE of 88.4% compared to only aggregating samples. This suggests that contrastive learning helps in learning distinctive features for separating different categories. However, the introduction of memory bank resulted in degraded segmentation accuracy of 86.7%. The study also found that when only conducting contrastive learning in the training stage, the pruning degraded the accuracy from 88.4% to 87.1%. However, when contrastive learning was introduced both in the training and fine-tuning stages, the accuracy was preserved at 87.8%.

Figure 7 visualizes the embedding distributions of Tiny-PPG learned with and without contrastive loss. Compared to the embedding distribution of non-contrastive learning, the distribution of contrastive learning is more uniform, and the intra-class samples are more gathered, demonstrating the core property of contrastive loss in intra-class attraction and inter-class separation. However, when the temperature coefficient increased, the embedding distribution became more aggregated, but some hard inter-class samples cannot be well separated. Overall, the intra-class compactness and inter-class separability brought by contrastive loss in this study was not significant, explaining the mild DICE improvement compared to non-contrastive learning.

**Real-time inference on MCU embedded system**. The PPG pulse sensor was tested on two locations: fingertip and earlobe. Data was collected for a 5-minute period for each setup. Motion artifacts were introduced by performing different movements, including tapping the sensor, nodding or swaying the head, as well as swinging the hand. The experimental setup and detection result of a 30-second segment are illustrated in Figure 8. We then annotated the data manually using the segmentation annotation tool in [24] and computed the DICE coefficient for each setup. The DICE coefficients for the fingertip and earlobe setups were 73.9% and 86.8% respectively. The real-time inference performance of the five lightweight models on the embedded system are summarized in Table 2, including the average inference time of thirty 30-sec signal segments and the total flash and RAM usage. It is worth noting that, unlike the segmentation models, which performed inference every 30 seconds, the 1-D CNN model with a 3-sec sliding window method as described in [17] conducted inference every 3 seconds. Therefore, the flash and RAM usage considerations were based on a 3-second signal length. To facilitate comparison with the segmentation models, the MCU inference time of the CNN model was multiplied by 10 in the table.

Table 2. The performance of Tiny-PPG with different contrastive learning configurations

| Model | Memory Bank Size | Contrastive Strategy | DICE |
|---|---|---|---|
| Tiny-PPG | N/A | N/A | 87.7% |
| Tiny-PPG (Contrastive train) | N/A | Artifact & Clean | **88.4%** |
| | N/A | Artifact only | 87.6% |
| | N/A | Clean only | 87.8% |
| | 1000 | Artifact & Clean | 86.7% |
| | 1000 | Clean only | 84.5% |
| | 1000 | Artifact only | 76.6% |
| *Pruned* Tiny-PPG (Only contrastive train) | N/A | Artifact & Clean | 87.1% |
| *Pruned* Tiny-PPG (Contrastive train & Contrastive fine-tune) | N/A | Artifact & Clean | **87.8%** |

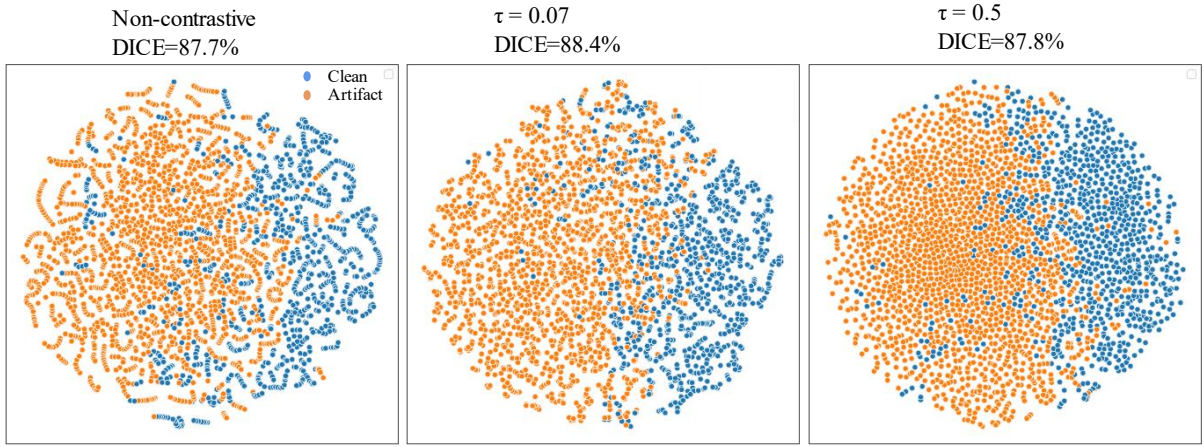

Figure 7. Visualization of embedding distributions learned *without* and *with* contrastive loss at different temperature coefficients.

Table 3. Performance comparison of five lightweight models deployed on the STM32 embedded device

| Models | MCU inference time (msec) | Total flash usage (kilobytes) | Total RAM usage (kilobytes) |
|---|---|---|---|
| 1D-CNN with 3-sec sliding window [17] | 838* | 209.56* | 51.356* |
| Baseline FCN | 478 | 127.32 | 267.624 |
| Baseline FCN-ASPP | 478 | 128.94 | 268.196 |
| Tiny-PPG | 328 | 89.064 | 196.260 |
| *Pruned* Tiny-PPG | 206 | 60.192 | 134.820 |

* The flash and RAM usage considerations were based on a 3-second signal length. The MCU inference time of the CNN model was multiplied by 10.

## 4. Discussion

Although convolutional neural networks such as Unet, Deeplab, and DeconvNet have been proved very effective in computer vision tasks such as medical image segmentation, their accuracy and model size has not been optimized for the application of PPG artifact detection. This study proposed a new lightweight fully convolutional network called Tiny-PPG, which adopted ASPP to provide multiscale features, and used computation and memory efficient depthwise separating convolutions, and was successfully implemented on a low-cost embedded system. Although previous studies have successfully implemented energy-efficient deep networks on-board for heart rate estimation based on PPG signal [35–37], it is important to consider the broader range of healthcare applications beyond this specific use case. These applications include cuffless blood pressure estimation [38], blood glucose estimation [39], blood oxygen saturation [4] and more. Since different applications may require different models, it is advantageous to have a separate step for detecting PPG artifacts. It achieved state-of-the-art accuracy in PPG artifact segmentation across a broad range of daily activities with only 20k model parameters after model pruning. Tiny-PPG was deployed in on a STM32 embedded system and was able to perform real-time detection of PPG artifacts.

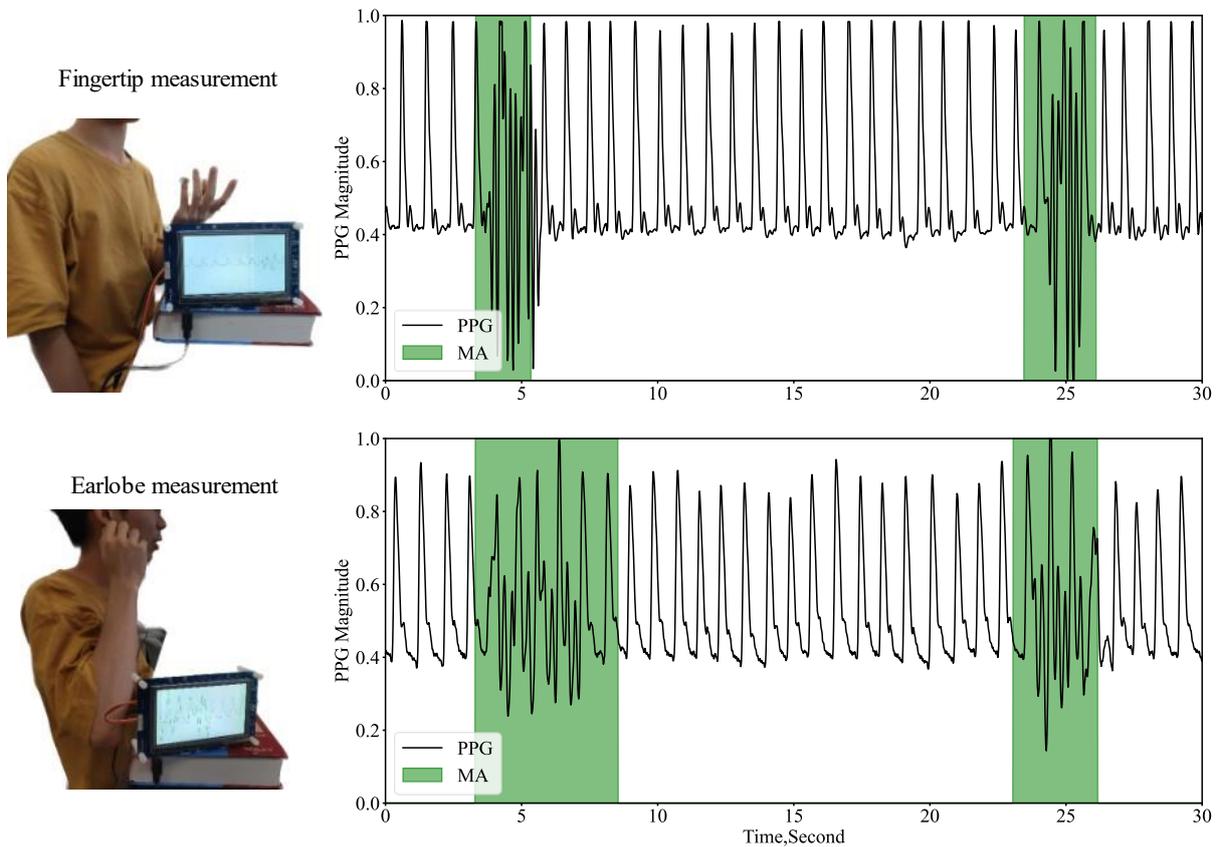

Figure 8. Real-time detection results of pruned Tiny-PPG on the STM32 embedded system with PPG signal recorded from the fingertip and earlobe of one subject.

In our preliminary results, it was found that the artifact segments were of various lengths and cannot be correctly detected by the baseline FCN. Therefore, we proposed to incorporate ASPP to provide a multi-scale receptive field, and the ablation experiment showed that ASPP module could slightly enhance the accuracy by 0.7%. In contrastive learning, it was found that the accuracy was not improved by only using the artifact or clean category for aggregating positive pairs. When using both the artifact and clean categories for aggregating positive pairs and separating negative pairs, the accuracy was notably improved, indicating that contrastive learning contributed to learning distinctive features for classification.

Incorporating the memory bank in contrastive learning generally degraded the accuracy possibly because the criteria for defining positive and negative pairs treated inter-class and intra-class samples as negative and positive pairs of the anchor sample, respectively. However, motion artifacts in PPG signals are inherently irregular, and their intrinsic similarities are relatively low, and PPG signals also vary greatly depending on age, gender, activity type and measurement location, etc. In other words, positive pairs are not necessarily similar with each other, and regularizing feature learning by aggregating them would lead to poor performance. There was still performance gain when only using 200 pairs in a mini-batch for the contrastive training, which is most possibly attributed to the distinctive features learned by separating positive and negative pairs. However, with the increase of number of positive and negative pairs when introducing the memory bank, the performance gain was compromised by aggregating unsimilar positive pairs. Therefore, more representative positive and negative pairs should be designed in the future to further leverage the advantage of contrastive learning.

Under the real-time test, the segmentation accuracy of Tiny-PPG was dependent on the measurement setup. The detection accuracy of the fingertip measurement setup was inferior to that of the earlobe setup, specifically 73.9% versus 86.8%. This discrepancy can be largely attributed to the significant difference in PPG morphology between the fingertip and the training dataset, in contrast to the earlobe. The PPG signals in the training dataset were acquired mainly from the wrist. We further found that the performance degraded significantly on one subject, the PPG morphology of whom was notably different from the training set. These results suggest that the generalization capability of Tiny-PPG to a more diverse range of subjects, especially subjects with different PPG morphologies, should be further improved in the future.

Previous studies have typically used segment lengths ranging from 5 to 60 seconds for signal quality evaluation. The choice of segment length in our study may be seen as a compromise between two factors: computation time and contextual information. Longer segments can provide more comprehensive contextual information for segmentation decisions, but they also require more computation time per segment. Additionally, we considered the ESC Guidelines for the management of atrial fibrillation, which recommend a minimum diagnostic episode duration of at least 30 seconds [40].

## 5. Conclusion

In conclusion, this study proposed a new lightweight deep neural network, called Tiny-PPG, for accurate and real-time PPG artifact detection on IoT edge devices. The model achieved state-of-the-art detection accuracy of 87.8% while having only 19,726 model parameters and was successfully deployed on an STM32 embedded system for real-time detection. By utilizing depth-wise separable convolution and atrous spatial pyramid pooling modules, as well as the contrastive loss, the model achieved a balance between detection accuracy and speed for resource-constrained embedded devices. Overall, the results suggest that Tiny-PPG provides an effective solution for PPG artifact detection in IoT-based wearable and smart health devices, and has the potential to significantly improve cardiovascular health monitoring in real-world settings. For reproducibility, the implementation of the proposed Tiny-PPG model in python and C languages has been published at https://github.com/SZTU-wearable/Tiny-PPG .

**Acknowledgement**

This work was supported by Guangdong Basic and Applied Basic Research Foundation [2021A1515110025],

**References**

[1]  S. Kekade, C.-H. Hseieh, M.M. Islam, S. Atique, A.M. Khalfan, Y.-C. Li, S.S. Abdul, The usefulness and actual use of wearable devices among the elderly population, Computer Methods and Programs in Biomedicine. 153 (2018) 137–159.

[2]  M. Panwar, A. Gautam, D. Biswas, A. Acharyya, PP-Net: A Deep Learning Framework for PPG-Based Blood Pressure and Heart Rate Estimation, IEEE Sensors J. 20 (2020) 10000–10011. https://doi.org/10.1109/JSEN.2020.2990864.

[3]  G. Thambiraj, U. Gandhi, U. Mangalanathan, V.J.M. Jose, M. Anand, Investigation on the effect of Womersley number, ECG and PPG features for cuff less blood pressure estimation using machine learning, Biomedical Signal Processing and Control. 60 (2020) 101942. https://doi.org/10.1016/j.bspc.2020.101942.

[4]  J.S. Hoffman, V.K. Viswanath, C. Tian, X. Ding, M.J. Thompson, E.C. Larson, S.N. Patel, E.J. Wang, Smartphone camera oximetry in an induced hypoxemia study, NPJ Digital Medicine. 5 (2022) 146.

[5]  P.H. Charlton, T. Bonnici, L. Tarassenko, D.A. Clifton, R. Beale, P.J. Watkinson, An assessment of algorithms to estimate respiratory rate from the electrocardiogram and photoplethysmogram, Physiol. Meas. 37 (2016) 610–626. https://doi.org/10.1088/0967-3334/37/4/610.

[6]  Z. Zhang, Z. Pi, B. Liu, TROIKA: A General Framework for Heart Rate Monitoring Using Wrist-Type Photoplethysmographic Signals During Intensive Physical Exercise, IEEE Trans. Biomed. Eng. 62 (2015) 522–531. https://doi.org/10.1109/TBME.2014.2359372.

[7]  J.A. Sukor, S.J. Redmond, N.H. Lovell, Signal quality measures for pulse oximetry through waveform morphology analysis, Physiol. Meas. 32 (2011) 369–384. https://doi.org/10.1088/0967-3334/32/3/008.

[8]  W. Karlen, K. Kobayashi, J.M. Ansermino, G.A. Dumont, Photoplethysmogram signal quality estimation using repeated Gaussian filters and cross-correlation, Physiol. Meas. 33 (2012) 1617–1629. https://doi.org/10.1088/0967-3334/33/10/1617.

[9]  Q. Li, G.D. Clifford, Dynamic time warping and machine learning for signal quality assessment of pulsatile signals, Physiological Measurement. 33 (2012) 1491.

[10]  M. Elgendi, Optimal signal quality index for photoplethysmogram signals, Bioengineering. 3 (2016) 21.

[11]  S. Vadrevu, M.S. Manikandan, Real-time PPG signal quality assessment system for improving battery

life and false alarms, IEEE Transactions on Circuits and Systems II: Express Briefs. 66 (2019) 1910–1914.

[12] T. Pereira, K. Gadhoumi, M. Ma, X. Liu, R. Xiao, R.A. Colorado, K.J. Keenan, K. Meisel, X. Hu, A supervised approach to robust photoplethysmography quality assessment, IEEE Journal of Biomedical and Health Informatics. 24 (2019) 649–657.

[13] R. Couceiro, P. Carvalho, R.P. Paiva, J. Henriques, J. Muehlsteff, Detection of motion artifact patterns in photoplethysmographic signals based on time and period domain analysis, CSASVM. 35 (2014) 2369–2388. https://doi.org/10.1088/0967-3334/35/12/2369.

[14] K. Li, S. Warren, B. Natarajan, Onboard Tagging for Real-Time Quality Assessment of Photoplethysmograms Acquired by a Wireless Reflectance Pulse Oximeter, IEEE Trans. Biomed. Circuits Syst. 6 (2012) 54–63. https://doi.org/10.1109/TBCAS.2011.2157822.

[15] A. Mahmoudzadeh, I. Azimi, A.M. Rahmani, P. Liljeberg, Lightweight photoplethysmography quality assessment for real-time IoT-based health monitoring using unsupervised anomaly detection, Procedia Computer Science. 184 (2021) 140–147.

[16] M. Feli, I. Azimi, A. Anzanpour, A.M. Rahmani, P. Liljeberg, An energy-efficient semi-supervised approach for on-device photoplethysmogram signal quality assessment, Smart Health. 28 (2023) 100390.

[17] C.-H. Goh, L.K. Tan, N.H. Lovell, S.-C. Ng, M.P. Tan, E. Lim, Robust PPG motion artifact detection using a 1-D convolution neural network, Computer Methods and Programs in Biomedicine. 196 (2020) 105596. https://doi.org/10.1016/j.cmpb.2020.105596.

[18] H. Shin, Deep convolutional neural network-based signal quality assessment for photoplethysmogram, Computers in Biology and Medicine. 145 (2022) 105430.

[19] J. Azar, A. Makhoul, R. Couturier, J. Demerjian, Deep recurrent neural network-based autoencoder for photoplethysmogram artifacts filtering, Computers & Electrical Engineering. 92 (2021) 107065. https://doi.org/10.1016/j.compeleceng.2021.107065.

[20] A.H.A. Zargari, S.A.H. Aqajari, H. Khodabandeh, A.M. Rahmani, F. Kurdahi, An Accurate Non-accelerometer-based PPG Motion Artifact Removal Technique using CycleGAN, ACM Trans. Comput. Healthcare. (2022) 3563949. https://doi.org/10.1145/3563949.

[21] J. Chen, K. Sun, Y. Sun, X. Li, Signal Quality Assessment of PPG Signals using STFT Time-Frequency Spectra and Deep Learning Approaches, in: 2021 43rd Annual International Conference of the IEEE Engineering in Medicine & Biology Society (EMBC), IEEE, Mexico, 2021: pp. 1153–1156. https://doi.org/10.1109/EMBC46164.2021.9630758.

[22] S.-H. Liu, R.-X. Li, J.-J. Wang, W. Chen, C.-H. Su, Classification of Photoplethysmographic Signal Quality with Deep Convolution Neural Networks for Accurate Measurement of Cardiac Stroke Volume, Applied Sciences. 10 (2020) 4612. https://doi.org/10.3390/app10134612.

[23] X. Liu, Q. Hu, H. Yuan, C. Yang, Motion Artifact Detection in PPG Signals Based on Gramian Angular Field and 2-D-CNN, in: 2020 13th International Congress on Image and Signal Processing, BioMedical Engineering and Informatics (CISP-BMEI), IEEE, Chengdu, China, 2020: pp. 743–747.

https://doi.org/10.1109/CISP-BMEI51763.2020.9263630.

[24] Z. Guo, C. Ding, X. Hu, C. Rudin, A supervised machine learning semantic segmentation approach for detecting artifacts in plethysmography signals from wearables, Physiol. Meas. 42 (2021) 125003. https://doi.org/10.1088/1361-6579/ac3b3d.

[25] D. Balemans, P. Reiter, J. Steckel, P. Hellinckx, Resource efficient AI: Exploring neural network pruning for task specialization, Internet of Things. 20 (2022) 100599. https://doi.org/10.1016/j.iot.2022.100599.

[26] G. Sivapalan, K.K. Nundy, S. Dev, B. Cardiff, D. John, ANNet: A lightweight neural network for ECG anomaly detection in IoT edge sensors, IEEE Transactions on Biomedical Circuits and Systems. 16 (2022) 24–35.

[27] P. Anbukarasu, S. Nanisetty, G. Tata, N. Ray, Tiny-HR: Towards an interpretable machine learning pipeline for heart rate estimation on edge devices, ArXiv Preprint ArXiv:2208.07981. (2022).

[28] A. Reiss, I. Indlekofer, P. Schmidt, K. Van Laerhoven, Deep PPG: Large-Scale Heart Rate Estimation with Convolutional Neural Networks, Sensors. 19 (2019) 3079. https://doi.org/10.3390/s19143079.

[29] A.G. Howard, M. Zhu, B. Chen, D. Kalenichenko, W. Wang, T. Weyand, M. Andreetto, H. Adam, Mobilenets: Efficient convolutional neural networks for mobile vision applications, ArXiv Preprint ArXiv:1704.04861. (2017).

[30] L.-C. Chen, G. Papandreou, F. Schroff, H. Adam, Rethinking atrous convolution for semantic image segmentation, ArXiv Preprint ArXiv:1706.05587. (2017).

[31] Z. Liu, J. Li, Z. Shen, G. Huang, S. Yan, C. Zhang, Learning Efficient Convolutional Networks through Network Slimming, in: 2017 IEEE International Conference on Computer Vision (ICCV), IEEE, Venice, 2017: pp. 2755–2763. https://doi.org/10.1109/ICCV.2017.298.

[32] L.-C. Chen, Y. Zhu, G. Papandreou, F. Schroff, H. Adam, Encoder-decoder with atrous separable convolution for semantic image segmentation, in: Proceedings of the European Conference on Computer Vision (ECCV), 2018: pp. 801–818.

[33] H. Noh, S. Hong, B. Han, Learning Deconvolution Network for Semantic Segmentation, in: 2015 IEEE International Conference on Computer Vision (ICCV), IEEE, Santiago, Chile, 2015: pp. 1520–1528. https://doi.org/10.1109/ICCV.2015.178.

[34] P.K. Lim, S.-C. Ng, N.H. Lovell, Y.P. Yu, M.P. Tan, D. McCombie, E. Lim, S.J. Redmond, Adaptive template matching of photoplethysmogram pulses to detect motion artefact, Physiological Measurement. 39 (2018) 105005.

[35] D. Biswas, L. Everson, M. Liu, M. Panwar, B.-E. Verhoef, S. Patki, C.H. Kim, A. Acharyya, C. Van Hoof, M. Konijnenburg, others, CorNET: Deep learning framework for PPG-based heart rate estimation and biometric identification in ambulant environment, IEEE Transactions on Biomedical Circuits and Systems. 13 (2019) 282–291.

[36] A. Burrello, D.J. Pagliari, M. Risso, S. Benatti, E. Macii, L. Benini, M. Poncino, Q-ppg: Energy-efficient ppg-based heart rate monitoring on wearable devices, IEEE Transactions on Biomedical Circuits and Systems. 15 (2021) 1196–1209.

[37] A. Burrello, D.J. Pagliari, P.M. Rapa, M. Semilia, M. Risso, T. Polonelli, M. Poncino, L. Benini, S. Benatti, Embedding temporal convolutional networks for energy-efficient ppg-based heart rate monitoring, ACM Transactions on Computing for Healthcare (HEALTH). 3 (2022) 1–25.

[38] M. Kim, H. Lee, K.-Y. Kim, K.-H. Kim, Deep Learning Model for Blood Pressure Estimation from PPG Signal, in: 2022 IEEE International Conference on Metrology for Extended Reality, Artificial Intelligence and Neural Engineering (MetroXRAINE), IEEE, 2022: pp. 1–5.

[39] S.S. Gupta, T.-H. Kwon, S. Hossain, K.-D. Kim, Towards non-invasive blood glucose measurement using machine learning: An all-purpose PPG system design, Biomedical Signal Processing and Control. 68 (2021) 102706.

[40] P. Kirchhof, S. Benussi, D. Kotecha, A. Ahlsson, D. Atar, B. Casadei, M. Castella, H.-C. Diener, H. Heidbuchel, J. Hendriks, others, 2016 ESC Guidelines for the management of atrial fibrillation developed in collaboration with EACTS, Kardiologia Polska (Polish Heart Journal). 74 (2016) 1359–1469.